\documentclass{llncs}

\usepackage{amsmath}
\usepackage{amssymb}

\usepackage{url}
\usepackage{epsfig}
\usepackage{multicol}
\usepackage{wrapfig}
\usepackage{boxedminipage}
\usepackage[Symbol]{upgreek}
\usepackage{colortbl}
\usepackage{listings}
\usepackage{rotating}

\newcommand{\updr}{\ensuremath{\rhd}}
\newcommand{\updl}{\ensuremath{\lhd}}

\newcommand{\fancy}[1]{{\cal #1}}

\newenvironment{denselist}{\begin{list}{\textbullet}{\setlength{\itemsep}{0em}\setlength{\parsep}{0em}}}{\end{list}}

\newcommand{\CBGF}{$\Upxi$BGF}

\lstdefinelanguage{pp}{%
  numbers=none,
  literate={SPACE}{{\ }}1 {EPSILON}{{$\varepsilon$}}1 {STRING}{{$\lambda$}}1
 {*}{{$^*$}}1 {+}{{$^+$}}1 {?}{{$?$}}1 {<}{{$\langle$}}1 {>}{{$\rangle$}}1
{PS-ZERO}{{$0$}}1 {PS-ONE}{{$1$}}1 {SUB-ONE}{{$_1$}}1 {SUB-N}{{$_n$}}1 {SUB-M}{{$_m$}}1 {ELLIPSIS}{{$\dots$}}1
{PS-PLUS}{{${+}$}}1 {PS-OPLUS}{{${\oplus}$}}1 {PS-STAR}{{${*}$}}1 {PS-OSTAR}{{${\otimes}$}}1,
  keywordstyle=\ttfamily\bfseries,
 morekeywords={unfold},
  columns=flexible,
  basicstyle=\ttfamily,
}
\usepackage[unicode,bookmarks=false,pdfstartview={FitH},%
            colorlinks,linkcolor=blue,urlcolor=blue,citecolor=blue,%
            pdfauthor={Dr. Vadim Zaytsev},%backref=page,%
            pdftitle={Guided Grammar Convergence}]{hyperref}
\begin{document}
	\author{Vadim Zaytsev, \href{mailto:vadim@grammarware.net}{\url{vadim@grammarware.net}}}
  \institute{SWAT/CWI and IvI/UvA, Amsterdam, The Netherlands}
	% \institute{Software Analysis \& Transformation Team (SWAT),\\
	% Centrum Wiskunde \& Informatica (CWI), The Netherlands}
	\title{Guided Grammar Convergence}
	\maketitle

\newcommand{\plus}{\ensuremath{\raisebox{0.1em}{$\scriptscriptstyle\mathord{+}$}}}
\renewcommand{\star}{\ensuremath{\raisebox{0.1em}{$\scriptstyle\mathord{*}$}}}
\newcommand{\opt}{\ensuremath{\mathord{?}}}

\begin{abstract}
	Relating formal grammars is a hard problem that balances between language equivalence (which is known
	to be undecidable) and grammar identity (which is trivial). In this paper, we investigate
	several milestones between those two extremes and propose a methodology for inconsistency management
	in grammar engineering. While conventional grammar convergence is a practical approach relying on human
	experts to encode differences as transformation steps, guided grammar convergence
	is a more narrowly applicable technique that infers such transformation steps automatically by
	normalising the grammars and establishing a structural equivalence relation between them.
	This allows us to perform a case study with automatically inferring bidirectional transformations between
	11 grammars (in a broad sense) of the same artificial functional language: parser specifications with
	different combinator libraries, definite clause grammars, concrete syntax definitions, algebraic data types,
	metamodels, XML schemata, object models.
\end{abstract}

%-------------------------------------------------------------------------
%-------------------------------------------------------------------------
\section{Introduction}

Modern grammar theory has shifted its focus from general purpose programming languages to a
broader scope of \emph{software languages} that comprise programming languages, domain specific
languages, markup languages, API libraries, interaction protocols, etc~\cite{Towards}. Such
software languages are specified by \emph{grammars in a broad sense} that still rely on the
familiar infrastructure of terminals, nonterminals and production rules, but specify general
commitment to grammatical structure found in software systems. In that sense, a type safe program
commits to a particular type system; a program that uses a library, commits to using its
exposed interface; an XML document commits to the structure defined by its schema --- failure to
commit in any of these cases would mean errors in interpretation of the language entity. These,
and many other, scenarios can be expressed and resolved in terms of grammar technology, but not
all structural commitments profit from grammatical approach (as the most remarkably problematic
ones we can note indentation policies and naming conventions).

One of the problems of multiple implementations of the same language, which is known for many years, is having an
abstract syntax definition and a concrete syntax definition~\cite{Wile1997}. Basically, the abstract syntax
defines the kind of entities that inhabit the language and must be handled by semantics specification. A concrete
syntax shows how to write down language entities and how to read them back. It is not uncommon for a programming
language to have several possible concrete syntaxes: for example, any binary operation may use prefix, infix or
postfix notation, without any changes to the language semantics. Indeed, we have seen infix dialects of postfix
Forth (Forthwrite, InfixForth) and prefix dialects of infix REBOL (Boron). For software languages, the problem is
broader: we can speak of one \emph{intended language} specification and a variety of abstract and concrete
syntaxes, data models, class dictionaries, metamodels, ontologies and similar contracts that conform to it.

% by Charles Moore and and Andrew
% Haley\footnote{\url{http://www.complang.tuwien.ac.at/anton/euroforth/ef08/papers/haley.pdf}}
% by Carl Sassenrath and Karl
% Robillard\footnote{\url{http://urlan.sourceforge.net/boron/}}

Our definition of the intended language relies on bidirectional
transformations~\cite{bxReport,DCMUI1998,Stevens07,Metasyntactically2012} and in particular on
their notation by Meertens \cite{DCMUI1998}, which we redefine here for the sake of completeness and clarity:

\vspace{-.5em}\begin{definition}\label{def:meertens}
	For a relation $R \subseteq S \times T$, a \textbf{semi-maintainer} is a function
	$\updr:S \times T \to T$, such that
	$\forall x\in S, \forall y \in T, \langle x, x \updr y \rangle \in R$, and
	$\forall x\in S, \forall y \in T, \langle x, y \rangle \in R \Rightarrow x \updr y = y$.
\end{definition}\vspace{-.5em}

The first property is called \emph{correctness} and ensures that the update caused by the semi-maintainer restores
the relation. The second property is \emph{hippocraticness} and states that an update has no effect (``does no
harm''), if the original pair is already in the relation~\cite{Stevens07}. Other properties of bidirectional
transformations such as \emph{undoability} are often unachievable. A \emph{maintainer} is a pair of
semi-maintainers $\updr$ and $\updl$. A \emph{bidirectional mapping} is a relation and its maintainer.

\vspace{-.5em}\begin{definition}\label{def:intended}
	A grammar $G$ conforms to the language \textbf{intended} by the master grammar $M$, if there
	exists a bidirectional mapping between instances of their languages.\vspace{-2em}
	\begin{align*}
			G \models L(M) \iff\:
			&\exists R \subseteq L(G) \times L(M)\\
			&\exists \updr:L(G)\times L(M) \to L(M)\\
			&\exists \updl:L(G)\times L(M) \to L(G)
	\end{align*}
\end{definition}\vspace{-.5em}

Naturally, for any grammar holds $G\models L(G)$.

For example, consider a concrete syntax $G_c$ of a programming language used by
programmers and an abstract syntax $M=G_a$ used by a software reengineering
tool. We would need \updr\ to produce abstract syntax trees from parse trees and
\updl\ to propagate changes done by a reengineering tool, back to parse trees.
If those can be constructed --- examples of algorithms have been
seen~\cite{TIF,ObjectGrammars2012,Wile1997}, --- then $G_c$ conforms to the
language intended by $G_a$. As another example, consider an object model used in
a tool that stores its objects in an external database (XML or relational): the
existence of a bidirectional mapping between entries (trees or tables) in the
database and the objects in memory, means that they represent the same intended
language, even though they use very different ways to describe it and one may be
a superlanguage of the other. For a more detailed formalisation and discussion
of bidirectional mappings and grammars, a reader is redirected
elsewhere~\cite{Bidirectionalisation2014,Parsing2014}.

{\footnotesize
\textbf{\emph{Roadmap.}}
In the following sections, we will briefly present the following milestones of relationships between languages:

\S\ref{S:identity}. \emph{Grammar identity}: structural equality of grammars

\S\ref{S:aware}. \emph{Nominal equivalence}: name-based equivalence of grammars

\S\ref{S:agnostic}. \emph{Structural equivalence}: name-agnostic footprint-matching equivalence

\S\ref{S:ANF}. \emph{Abstract normalisation}: structural equivalence of normalised grammars

Then, \S\ref{S:discuss} summarises the proposed method and discusses its evaluation.

Finally, \S\ref{S:conclusion} concludes the paper by establishing context and contributions.
}
\section{Grammar identity}\label{S:identity}

Let us assume that grammars are traditionally defined as quadruples $G=\langle\fancy{N},{\cal
T},\fancy{P},\fancy{S}\rangle$ where their elements are respectively the sets of nonterminal
symbols, terminal symbols, production rules and starting nonterminal symbols.

\vspace{-.5em}\begin{definition}\label{def:equality}
	Grammars $G$ and $G'$ are \textbf{identical}, if and only if all their components are identical:
	$
		G = G' \iff \fancy{N}=\fancy{N}' \wedge \fancy{T}=\fancy{T}' \wedge \fancy{P}=\fancy{P}' \wedge
		\fancy{S}=\fancy{S}'
	$.
\end{definition}\vspace{-.5em}

The definition is trivial, and in practice is commonly weakened somehow. For
example, many metalanguages allow the right hand sides of rules from $\fancy{P}$
to contain disjunction (inner choice), which is known to be commutative, so it
is natural to disregard the order of disjunctive clauses when comparing
grammars: \texttt{gdt}, the ``grammar diff tool'' used in convergence case
studies \cite{Convergence2009,SLPS} implements that. However, many grammar
manipulation technologies such as PEG~\cite{PEG} or TXL~\cite{DeanCMS02}, use
ordered choices, so this optimisation can be perceived as premature. For this
reason, we will explicitly abandon disjunction in later sections.

\section{Nominal equivalence}\label{S:aware}

Since identity can be seen as a trivial bijection, a disciplined weakening of \autoref{def:equality}
that works across all grammars in a broad sense, is this:

\vspace{-.5em}\begin{definition}\label{def:nominal}
	Grammars $G$ and $G'$ are \textbf{nominally equivalent}, if there is a bijection $\beta$
	between their production rules:\vspace{-.5em}
	\begin{gather*}
		G \approxeq G' \iff \fancy{N}=\fancy{N}' \wedge \fancy{T}=\fancy{T}' \wedge \fancy{S}=\fancy{S}' \wedge
		\exists \beta:\fancy{P}\to\fancy{P}',\\
		\forall q\in\fancy{P}', \exists p\in\fancy{P}, q=\beta(p);\qquad
		% \forall p\in\fancy{P}, \beta(p)\in\fancy{P}';
		\forall p_1,p_2\in\fancy{P}, \beta(p_1)=\beta(p_2)\Rightarrow p_1=p_2
	\end{gather*}
\end{definition}\vspace{-.5em}

Algorithms that are used to construct $\beta$ can be different. For example, in Popart the
metalanguage is designed in such a way that it contains enough information to generate both abstract
and concrete syntaxes~\cite{Wile1997}. In TIF-grammars, a concrete syntax specification is annotated
with directions on which nodes need to be folded/unfolded or removed when constructing an abstract
syntax tree (AST)~\cite{TIF}. In Rascal, the \texttt{implode} function that maps parse trees to ASTs
uses names of nonterminals and subexpressions to direct the automatic construction of $\beta$: for
example, an optional nonterminal occurrence can be mapped to a string,
 % (the absence signalled by an empty string, in the presence its contents are cast to string), or to
a list
% (of zero or one element), or even to a
or a Boolean;
 % (true for presence, false for absence);
a Kleene star can be imploded to either a list or a
set, etc~\cite{Rascal}. Similar techniques can be spotted in OOP~\cite{DigCMJ06}, in
MDE~\cite{FalleriHLN08}, in data binding frameworks~\cite{JSR31}, etc.

% data binding?

Once the bijective $\beta$ is agreed on the level of grammars, we need to construct a coupled maintainer on the
language instance level. If it cannot be constructed, then $\beta$ is useless for us. However, there are many
cases when the maintainer can be constructed to be bidirectional ($R$ from \autoref{def:intended} can be partial
and \updr\ and/or \updl\ can be only injective but not surjective). An example of that is matching different
representations of lists/sets: ``one or more'' and ``zero or more'' Kleene repetitions are commonly used in
syntactic notations, but one can always be bidirectionally matched to the other. In our prototype implementation,
we disregard unreachable nonterminals, treat built-in and user-defined nonterminals equally, allow sequence
element permutations and desugar metasyntactic constructs like ``separator lists'', as well as match different
kinds of lists/sets. Such strategy set was chosen to be valid for nominal matching techniques, but also useful for
the following sections.
\section{Structural equivalence}\label{S:agnostic}

In order to reuse the methods from the nominal equivalence approach in the case when nonterminal names do not
match, we need to construct an additional mapping between the nonterminals, and use that instead of nominal
identity. We shall refer to this mapping as \emph{nominal resolution}. To construct it, we generalise
permutations and construct \emph{signatures} that express general structure of a production rule. Such signatures
depend heavily on the expressiveness of the chosen metalanguage (in particular, when the formal grammar is
defined with just terminals and nonterminals, their use is severely limited), but most common metalanguages
contain enough functionality to allow signatures to work.

\vspace{-.5em}\noindent\begin{minipage}{.5\textwidth}
	\begin{definition}\label{def:footprint}
		A \textbf{footprint} $\pi_n(x)$ of a nonterminal $n$ in an expression $x$ is defined as a multiset of presence indicators.
	\end{definition}
	\begin{definition}\label{def:fp-equiv}
		Two footprints are \textbf{equivalent}, if they are equal modulo repetition kinds:
		$ \pi\approx\xi \Longleftrightarrow 
			\pi = \xi \vee
			\pi' = \xi',$
		where $\zeta'$ is $\zeta$ with all $\plus$ elements replaced by $\star$ elements.
	\end{definition}
\end{minipage}
\begin{minipage}{.5\textwidth}
	{\footnotesize
	$$ \pi_n(x) \stackrel{def}{=}
	\begin{cases}
		\{1\}	& \text{if } x=n\\
		\{?\}	& \text{if } x=n?\\
		\{+\}	& \text{if } x=n^{+}\\
		\{*\}	& \text{if } x=n^{*}\\
		\pi_n(y)& \text{if } x={}_{name:}y\\
		% \bigcup\limits_{e\in L} \pi_n(e)	& \text{if } x=\mathrm{seq}(L)\\
		\pi_n(e_1)\cup\pi_n(e_2)& \text{if } x=e_1 e_2\\
		\varnothing	& \text{otherwise}
	\end{cases} $$}
	
	\phantom{!}
	
	\phantom{!}
\end{minipage}\vspace{-.5em}

Note how disjunction is missing from the definition of a footprint. We will see later how it can be removed from
any grammar by factoring and folding. The identity of footprints follows the standard definition of identity of
multisets (which naturally subsumes abstraction over permutations). We also define equivalence of them that
generalises the treatment of lists discussed in the previous section.
% Naturally, $\forall \pi, \forall\xi,\: \pi=\xi \Rightarrow \pi\approx\xi$. 
Footprints together form a signature.

\vspace{-.5em}\begin{definition}\label{def:signature}
	A prodsig, or a \textbf{signature} of a production rule $p = (m ::= x)$ is defined as a set of tuples with
	nonterminals used in its right hand side and their footprints:
	% $$\sigma(\p(a,m,x)) = \{\langle n, \pi_n(x) \rangle\: |\: n \in \textrm{usedNs}(x)\}$$
	$\qquad\sigma(m ::= x) = \{\langle n, \pi_n(x) \rangle\: |\: n \in \fancy{N}, \pi_n(x)\neq\varnothing\}$.
	% Just like in the previous section, for the purpose of constructing a prodsig, values (built-in syntactic
	% categories like strings or integers) are treated as nonterminals and complex subexpressions can be seen as
	% pseudo-nonterminals introduced by transparent folding transformations.
\end{definition}\vspace{-.5em}

For example, the prodsig of a production rule $P ::= F^{+}$ is $\{\langle F, \{+\}\rangle\}$ and the prodsig of
$F ::= S S^{*} E$ is $\{\langle E, \{1\}\rangle, \langle S, \{1,{*}\}\rangle\}$.

\vspace{-.5em}\begin{definition}\label{def:weak-equiv}
	Two production rules are \textbf{prodsig-equivalent},
	if and only if there is an equivalent %unique
	match between tuple ranges of their signatures:\\
	$\phantom{there is a unique}p \Bumpeq q \iff
		\forall\langle n,\pi\rangle\in\sigma(p),\:
		\exists \langle m,\xi\rangle\in\sigma(q),\:
		\pi\approx\xi$
\end{definition}\vspace{-.5em}

% non-commutativity

Consider a simple case of exactly one production rule taken from each of the grammars: $p_m$ from
the master grammar and $p_s$ from the servant grammar. Suppose that the left hand sides of them
are assumed to match, and we want to see if the right hand sides are matched nominally as well,
and whether they deliver any new information with respect to nominal resolution. When prodsigs
$\sigma(e_m)$ and $\sigma(e_s)$ are constructed, we have effectively built relations that bind
nonterminal names to their occurrences. By subsequently matching the \emph{ranges} of them with
either strong or weak prodsig-equivalence, we can infer nominal matching of the nonterminals by
matching the \emph{domains} of the relations.

%And for each two prodsig-equivalent production rules it becomes trivial to define the nominal matching relationship:

% \begin{definition}\label{def:nominal-strong}
% 	For any two strongly prodsig-equivalent production rules $p$ and $q$, $p\bumpeq q$, a
% 	\textbf{nominal resolution}	relationship has the form of:\\
% 	$p \diamond q = \sigma(p) \circ \overline{\sigma(q)} $,
% 	where $\rho_1 \circ \rho_2$ is a composition of two relations in the sense of $\rho_1 \circ \rho_2 \stackrel{\text{def}}{=} \{\langle a,c\rangle | \langle a,b\rangle\in\rho_1, \langle b,c\rangle\in\rho_2\}$, and $\overline{\rho}$ is the inverse of a relationship in the sense of $\bar{\rho} \stackrel{\text{def}}{=} \{\langle a,b\rangle | \langle b,a\rangle\in\rho\}$.
% \end{definition}

\vspace{-.5em}\begin{definition}\label{def:nominal-weak}
	For any two prodsig-equivalent production rules $p$ and $q$, $p\Bumpeq q$,
	% if we use $\omega$ to denote unmatched nonterminals, th
	there is (at least one)
	nominal	resolution relationship $p \diamond q$ that satisfies:
\vspace{-.5em}\begin{align*}
	\forall \langle a,b\rangle \in p \diamond q:\quad
	&a = \omega \vee b = \omega \:\vee
	\exists \pi, \exists \xi,
	\pi\approx\xi,
	\langle a,\pi\rangle\in\sigma(p),
	\langle b,\xi\rangle\in\sigma(q)\\
	% \textrm{\emph{ and }}
	\forall \langle c,d\rangle \in p \diamond q:\quad
	&
	% \forall \langle a,b\rangle \in p \diamond q,
	% \forall \langle c,d\rangle \in p \diamond q:
	a = c \neq \omega \Rightarrow b = d,
	\textrm{where $\omega$ denotes unmatched nonterminals.}
\end{align*}
% 
% 	\begin{eqnarray*}
% 	p \:\Diamond\: q &=& \left\{\langle a,b\rangle | \langle a,\pi\rangle\in\sigma(p), \langle b,\xi\rangle\in\sigma(q), \pi\approx\xi\right\}\\
% 	&\cup& \{\langle\omega,c\rangle | \langle c,\xi\rangle\in\sigma(q), \forall \pi\approx\xi,\nexists a:\langle a,\pi\rangle\in\sigma(p)\}
% 	\end{eqnarray*}
\end{definition}\vspace{-.5em}

For two arbitrarily provided grammars (presumably of the same intended software language, but not
necessarily admitting any kind of equivalence), we cannot claim the existence of only one nominal
resolution that works across all their production rules, but we can attempt to construct a
minimum possible one:
% but the algorithm that searches for it,
% can be trivially programmed by taking its definition.

\vspace{-.5em}\begin{definition}\label{def:nominal-final}
	Given two grammars $G_1$ and $G_2$, a nominal resolution between them is a
	relation between their nonterminals	$\Diamond\in\fancy{N}_1\times\fancy{N}_2$ such that
	$\forall p_i\in\fancy{P}_1$, if $\exists q_j\in\fancy{P}_2, p_i\Bumpeq q_j$, then
	$\exists \diamond_{ij}\subset\Diamond$, such that $p_i \diamond_{ij} q_j$.
	% 
	% A nominal resolution between two grammars $G_1$ and $G_2$ is a relation between their nonterminals
	% $\Diamond\in\mathbb{N}_1\times\mathbb{N}_2$ such that
	% $\forall p_i\in\mathbb{P}_1, \exists q_j\in\mathbb{P}_2, p_i\Bumpeq q_j$, then
	% $\exists \diamond_{ij}\subset\Diamond$, such that $p_i \diamond_{ij} q_j$.
\end{definition}\vspace{-.5em}

In our case study, we have used various definitions of the same toy functional language FL taken
from~\cite{Convergence2009}. For instance, some grammars were extracted from object models by analysing Java
code. Since the Java implementation of FL used \texttt{List<Expr>} to represent lists, the production rules for
function declaration and function call assumed zero or more arguments, while the master grammar assumed one or
more. Hence, production rules $F_1 ::= S_1 S_1^{*} E_1$ and $E_1 ::= S_1 E_1^{*}$ were matched with their
prodsig-equivalent counterparts $F_2 ::= S_2 S_2^{+} E_2$ and $E_2 ::= S_2 E_2^{+}$.
All the various grammars of FL, their prodsigs and nominal matching reports are exposed for
inspection in the full report on the case study~\cite{Guided2012}.

% \newpage
\section{Abstract normalisation}
% \section{Abstract Normal Form}
\label{S:ANF}

In order to apply the methodology based on nonterminal footprints, production signatures and their equivalence
relations, we need the input grammars to comply with some assumptions that have been left informal so far. In
particular, we can foresee possible
problems with names/labels for production rules and subexpressions, terminal symbols (often not a part of the
abstract syntax), disjunction (inner choices, also non-factored), separator lists and other metasyntactic sugar,
non-connected nonterminal call graph, inconsistent style of production rules, etc.
If by ${\cal P}_n\subset{\cal P}$ we denote
the subset of production rules concerning one particular nonterminal:
% \vspace{-.5em}
$
	{\cal P}_n=\{p\in{\cal P} \: | \: p = n ::= \alpha,\: \alpha\in({\cal N\cup T})^*\},
	% \qquad \text{where }n\in{\cal N}
$
then we can define the Abstract Normal Form as follows:

\begin{definition}\label{def:anf}
	% Given two disjoint sets of nonterminals, ${\cal N}_d$ and ${\cal N}_\bot$, 
	A grammar $G=\langle{\cal N},{\cal T},{\cal P},{\cal S}\rangle$, where ${\cal T}=\varnothing$ and ${\cal S}=\{s\}$, is said to be in Abstract Normal Form, if and only if:
\begin{denselist}
	% \item Terminal symbols are not present: ${\cal T}=\varnothing$
	% \item ${\cal S}$ consists of only one root element: ${\cal S}=\{s\}$
	\item ${\cal N}$ is decomposable to disjoint sets, such that
		${\cal N}={\cal N}_{+}\cup{\cal N}_{-}\cup{\cal N}_\bot$
	\item One of them is not empty and includes the root:
		$s\in{\cal N}_{+}\cup{\cal N}_{-}$
	\item Nonterminals from one subset are undefined:
		$n\in{\cal N}_\bot \Rightarrow {\cal P}_n=\varnothing$
	\item Nonterminals from one other subset are defined with exactly one rule:\\
	% \begin{equation}\label{eq:onerule}
	$
		n\in{\cal N}_{-} \Rightarrow
			|{\cal P}_n|=1,\:
			{\cal P}_n = \{n::=\alpha\},\:
			% p= n\to\alpha,\:
			\alpha\in{\cal N}^{+} %\setminus {\cal N}
	$
	\item Nonterminals from the other subset are defined with chain rules:\\
		$n\in{\cal N}_{+} \Rightarrow
			\forall p\in{\cal P}_n,\:
			p= (n::= x),\:
			x\in{\cal N}$
\end{denselist}	
\end{definition}

% all these issues can be addressed, and 

In fact, any grammar can be rewritten to assume this form: in our prototype
implementation, this is done by programming a \emph{grammar
mutation}~\cite{Metasyntactically2012,SLEIR2014}. (A grammar mutation is a
general intentional grammar change that cannot be expressed independently of the
grammar to which it will be applied. Thus, if ``rename'' is a parametric grammar
transformation operator, then ``rename A to B'' is a transformation, but
``rename all nonterminals to uppercase'' is a mutation that is equivalent to
transformations like ``rename a to A'' and ``rename b to B'' depending on the
input grammar). Our prototype, \texttt{normal::ANF}, is a metaprogram in
Rascal~\cite{Rascal} that is available for inspection as open
source~\cite{SLPS}. It is in fact a superposition of mutations that address the
items from the definition individually: remove labels, desugar separator lists,
fold/unfold chain production rules, etc.

All the rewritings performed by transforming a grammar to its ANF, are assumed
to be monadic in the sense of not only normalising the grammar, but also
yielding a bidirectional grammar transformation chain which execution would
normalise the grammar. (In our implementation, these steps are specified in the
\CBGF\ language, primarily because no other bidirectional grammar transformation
operator suite exists~\cite{SLEIR2014}). The bidirectional grammar
transformation chain can then be coupled to the bidirectional mapping between
language instances from \autoref{def:intended}, with the methodology described
by \cite{Metasyntactically2012}. This is required for traceability: the
conversion to ANF is one of the steps to achieve automated convergence, not a
one-way preprocessing.

% \section{Concluding remarks}
\section{Discussion}\label{S:discuss}

To summarise, grammar convergence is a technique of relating different grammars in a broad sense of the same
intended software language~\cite{Convergence2009}. It relies on the transformations being programmed by an
experienced grammar engineer: even beside the required expertise, the process is not incremental --- the
transformation steps need to be considered carefully and constructed for each new grammar added to the mix. With
the definitions from previous sections, we have described the process of \emph{guided grammar convergence}, where
the master grammar of the intended language is constructed once, and the transformations are inferred for any
directly available grammars as well for the ones possibly added in the future. The process works as follows.

\begin{denselist}
	\item Extract pure grammatical knowledge from the grammar source.
	\item Use grammar mutations to preprocess your grammars, if necessary.
	\item Normalise the grammar by removing all problematic/ambiguous constructs.
	\item Start by matching the roots of the connected normalised grammar.
	\item Match multiple production rules by prodsig-equivalence; infer new nominal matches by matching equivalent
	prodsigs. Repeat for all nonterminals.
	% \item Establish nominal resolution by repeating last two steps for newly matched nonterminals
	\item If several matches are possible, explore all and fallback in case of failure.
		If global nominal resolution scheme was impossible to infer, fail.
	\item Resolve structural differences in the production rules that matched nominally.
\end{denselist}

To evaluate the method of guided grammar convergence, we have applied it to a case study of \textbf{11} different
grammars of the same intended functional language that was defined and used earlier in order to demonstrate the
original grammar convergence method that converged \textbf{5} of these grammars. The following grammar sources
were used (all of them are available in the repository of SLPS~\cite{SLPS} together with their evolution history
and authorship attribution):

\begin{description}
\item[adt:] an algebraic data type\footnote{\url{http://tutor.rascal-mpl.org/Rascal/Declarations/AlgebraicDataType/AlgebraicDataType.html}.} in Rascal~\cite{Rascal};
\item[antlr:] a parser description in the input language of ANTLR~\cite{ANTLR}, with
semantic actions (in Java) intertwined with EBNF-like productions;
\item[dcg:] a logic program written in the style of definite clause grammars~\cite{DCG};
% \item[ecore:] an Ecore model~\cite{MOF}, created manually in Eclipse modeling framework~\cite{EMF}
% 	and represented in XMI;
\item[emf:] an %alternative 
	Ecore model, automatically generated by Eclipse~\cite{EMF} from the XML Schema of the
	\textbf{xsd} source;
\item[jaxb:] an object model obtained by a data binding framework,
generated automatically by JAXB~\cite{JSR31} from the XML schema for FL;
\item[om:] a hand-crafted object model (Java classes) for the abstract syntax of FL;
\item[python:] a parser specification in a scripting language, using the PyParsing library~\cite{McGuire2007};
\item[rascal:] a concrete syntax specification in the metaprogramming language of Rascal language workbench \cite{Rascal};
\item[sdf:] a concrete syntax definition in the notation of SDF~\cite{ASFSDF-Klint} with scannerless
generalized LR parsing as parsing model.
\item[txl:] a concrete syntax definition in the notation of TXL (Turing eXtender Language) transformational framework \cite{DeanCMS02}, which, unlike SDF, uses a combination of pattern matching and term rewriting).
\item[xsd:] an XML schema~\cite{W3C-XSD} for the abstract syntax of FL.
\end{description}

The complete case study is too big to be presented here, interested readers are redirected to a 40+ page long
report~\cite{Guided2012}, containing all production rules, signatures, matchings and transformations. The case
study was successful: on average ANF was achieved after 20--30 transformation steps, nominal resolution took up to
9 (proportional to the number of nonterminals) and structural resolution needed 0--5 more steps, after which all
11 grammars were converged. ANTLR, DCG and PyParsing used layered definitions and therefore were the only three
grammars to require mutation (another 2--6 steps). The case study is available for investigation and replication
both in the form of Rascal metaprograms at \url{http://github.com/grammarware/slps}~\cite{SLPS} (the main
algorithm is located in the \texttt{converge::Guided} module which can be observed and modified at
\href{http://github.com/grammarware/slps/blob/master/shared/rascal/src/converge/Guided.rsc}{\texttt{shared/rascal/src/converge/Guided.rsc}})
and as a PDF report with all grammars and transformations pretty-printed automatically~\cite{Guided2012}.

Guided grammar convergence is a methodology stemming from the grammarware ``technological
space''~\cite{KurtevBA02}. When looking for similar techniques in other spaces (engaging in
``space travel''), the obvious candidates are schema matching and data
integration in the field of data modeling and databases~\cite{RahmB01}; comparison of UML models
or metamodels in the context of model-driven engineering~\cite{FalleriHLN08,XingS06a}; model
weaving for product line software development~\cite{DelFabro2007,Voelter2007}; computation of
refactorings from different OO program versions~\cite{DigCMJ06,OKeeffe08}; etc. For example,
Del Fabro and Valduriez~\cite{DelFabro2007} utilised metamodel properties for automatically producing weaving models. The
core difference is that (meta)model weaving ultimately aims at incorporating all the changes into
the resulting (meta)model, while guided grammar convergence also makes complete sense when some
changes in the details are disregarded. The lowest limit in this process is needed (otherwise
additional claims on the minimality of inferred transformations are required), and we specify
this lowest limit as the master grammar. Another difference is that model weaving rarely involves
a number of models bigger than two, and even our little case study of guided convergence had 10+
grammars in it. In general, prodsig-based matching is more lightweight than those methods, since it in fact
compares straightforwardly structured prodsigs, thus easily wins in performance and implementability but
loses in applicability to complex scenarios.
\section{Conclusion}\label{S:conclusion}

We knew that language equivalence is undecidable and that grammar identity is trivial. In this
paper, we have attempted to reach a useful level of reasoning about language relationships by
departing from grammar identity as the ``easy'' side of the spectrum. This was done in the scope
of grammar convergence, when several implementations of the same software language are inspected
for compatibility.

A definition of an \emph{intended language} was provided (\autoref{def:intended}) based on a bidirectional
transformation between language entities. Then we revisited existing and possible techniques of structural
matching that assumed nominal identity %$\fancy{N}_1=\fancy{N}_2$
(\autoref{def:equality}). In order to automatically infer nominal matching,
%$\Diamond\in\fancy{N}_1\times\fancy{N}_2$,
we introduced nonterminal footprints (\autoref{def:footprint}), production signatures (\autoref{def:signature})
and various degrees of equivalence among them. An extensive normalisation scheme (\autoref{def:anf}) was proposed
to transform any given grammar into the form most suitable for nominal and then structural matching. It has been
explained that when such normalisation is not enough, a more targeted yet still automated approach is needed with
grammar mutation strategies making the method robust with respect to different grammar design decisions, such as
the use of layers instead of priorities or recursion instead of iteration. Just as all other parts of the proposed
process, these mutations operate automatically and do not require human intervention.

A case study was used to evaluate the proposed method of guided grammar convergence. The
experiment concerned several implementations of a simple functional language in ANTLR, DCG,
Ecore, Java, Python, Rascal, SDF, TXL, XML Schema. The diversity in language processing
frameworks --- metaprogramming languages, declarative specifications, syntax definitions,
algebraic data types, parsing libraries, transformation frameworks, software models, parser
definitions --- was intentional and aimed at stressing the definition of the intended language
and the guided convergence method. Casting all grammars from our case study to ANF allowed us to
make inference quicker and with less obstacles, as well as to explain the process more clearly.

All artifacts discussed on the pages of this paper, are transparently available to the public
through a GitHub repository~\cite{SLPS}. For each of the sources of the case study, one could
inspect the original file, the extracted grammar, the extractor itself, the mutations that have
been derived and applied, the normalisations to ANF, the normalised grammar, the nominal
resolution and reasons for each match, as well as the structural resolution steps. One could also
investigate the implementation of the method of guided grammar convergence, the algorithm for
calculating prodsigs and the process of convergence. Supplementary material contains 40+ pages of
the full report, also generated by our prototype~\cite{Guided2012}.

On the practical side, guided grammar convergence provides a balanced method of grammar
manipulation, positioned right between unstructured inline editing (which makes grammar
development very much like software development but lacks important properties such as
traceability and reproducibility) and strictly exogenous functional transformation (which
requires substantially more effort but is robust, repeatable and exposes the intended semantics). Its
future role can be seen as a support for \textbf{grammar product lines} that allows both steady
adaptation plans for deriving secondary artifacts from the reference grammar, and occasional
inline editing of the derived artifacts with subsequent automated restoration of the adaptation
scripts. This is a contribution to the field of engineering discipline for
grammarware~\cite{Towards}.

{
% \footnotesize
\bibliographystyle{abbrv}
\bibliography{paper}

\begin{thebibliography}{10}

\bibitem{bxReport}
K.~Czarnecki, J.~Foster, Z.~Hu, R.~L{\" a}mmel, A.~Sch{\" u}rr, and
  J.~Terwilliger.
\newblock {Bidirectional Transformations: A Cross-Discipline Perspective}.
\newblock In {\em TPMT}, volume 5563 of {\em LNCS}, pages 260--283. Springer,
  2009.

\bibitem{DeanCMS02}
T.~R. Dean, J.~R. Cordy, A.~J. Malton, and K.~A. Schneider.
\newblock {Grammar Programming in TXL}.
\newblock In {\em {SCAM}}. IEEE, 2002.

\bibitem{DelFabro2007}
M.~D. Del~Fabro and P.~Valduriez.
\newblock {Semi-Automatic Model Integration Using Matching Transformations and
  Weaving Models}.
\newblock In {\em SAC}, pages 963--970, 2007.

\bibitem{DigCMJ06}
D.~Dig, C.~Comertoglu, D.~Marinov, and R.~Johnson.
\newblock {Automated Detection of Refactorings in Evolving Components}.
\newblock In {\em {ECOOP}}, volume 4067 of {\em LNCS}, pages 404--428.
  Springer, 2006.

\bibitem{EMF}
{Eclipse}.
\newblock {Eclipse Modeling Framework Project (EMF 2.4)}, 2008.
\newblock \url{http://www.eclipse.org/modeling/emf}.

\bibitem{FalleriHLN08}
J.-R. Falleri, M.~Huchard, M.~Lafourcade, and C.~Nebut.
\newblock {Metamodel Matching for Automatic Model Transformation Generation}.
\newblock In {\em {MoDELS}}, volume 5301 of {\em LNCS}, pages 326--340.
  Springer, 2008.

\bibitem{JSR31}
J.~Fialli and S.~Vajjhala.
\newblock {\em {Java Specification Request 31: XML Data Binding
  Specification}}, 1999.

\bibitem{PEG}
B.~Ford.
\newblock {Parsing Expression Grammars: a Recognition-Based Syntactic
  Foundation}.
\newblock In {\em POPL}, January 2004.

\bibitem{W3C-XSD}
S.~Gao, C.~M. Sperberg-McQueen, and H.~S. Thompson.
\newblock {W3C XML Schema Definition Language (XSD) 1.1}.
\newblock {\em {W3C Recommendation}}, Apr. 2012.

\bibitem{TIF}
A.~Johnstone and E.~Scott.
\newblock {Tear-Insert-Fold grammars}.
\newblock In {\em LDTA}, pages 6:1--6:8. ACM, 2010.

\bibitem{ASFSDF-Klint}
P.~Klint.
\newblock {Meta-Environment for Generating Program- ming Environments}.
\newblock {\em TOSEM}, 2(2):176--201, 1993.

\bibitem{Towards}
P.~Klint, R.~L{\"a}mmel, and C.~Verhoef.
\newblock {Toward an Engineering Discipline for Grammarware}.
\newblock {\em TOSEM}, 14(3):331--380, 2005.

\bibitem{Rascal}
P.~Klint, T.~van~der Storm, and J.~Vinju.
\newblock {EASY Meta-programming with Rascal}.
\newblock In {\em {GTTSE}}, volume 6491 of {\em LNCS}, pages 222--289.
  Springer, January 2011.

\bibitem{KurtevBA02}
I.~Kurtev, J.~B\'ezivin, and M.~Ak{\c s}it.
\newblock {Technological Spaces: an Initial Appraisal}.
\newblock In {\em CoopIS, DOA'2002, Industrial track}, 2002.

\bibitem{Convergence2009}
R.~L{\"a}mmel and V.~Zaytsev.
\newblock {An Introduction to Grammar Convergence}.
\newblock In {\em {iFM}}, volume 5423 of {\em LNCS}, pages 246--260. Springer,
  February 2009.

\bibitem{McGuire2007}
P.~McGuire.
\newblock {\em {Getting Started with Pyparsing}}.
\newblock O'Reilly, first edition, 2007.

\bibitem{DCMUI1998}
L.~Meertens.
\newblock {Designing Constraint Maintainers for User Interaction}.
\newblock Manuscript, June 1998.

\bibitem{OKeeffe08}
M.~O'Keeffe and M.~O. Cinn\'{e}ide.
\newblock {Search-Based Refactoring: an Empirical Study}.
\newblock {\em JSME}, 20(5):345--364, 2008.

\bibitem{ANTLR}
T.~Parr.
\newblock {ANTLR---ANother Tool for Language Recognition}, 2008.

\bibitem{DCG}
F.~Pereira and D.~Warren.
\newblock {Definite Clause Grammars for Language Analysis}.
\newblock In {\em {Readings in Natural Language Processing}}, pages 101--124.
  MKP, 1986.

\bibitem{RahmB01}
E.~Rahm and P.~A. Bernstein.
\newblock {A Survey of Approaches to Automatic Schema Matching}.
\newblock {\em The VLDB Journal}, 10:334--350, December 2001.

\bibitem{Stevens07}
P.~Stevens.
\newblock {Bidirectional Model Transformations in QVT: Semantic Issues and Open
  Questions}.
\newblock In {\em {MODELS}}, volume 4735 of {\em LNCS}, pages 1--15. Springer,
  2007.

\bibitem{ObjectGrammars2012}
T.~v.~d. Storm, W.~R. Cook, and A.~Loh.
\newblock {Object Grammars: Compositional \& Bidirectional Mapping Between Text
  and Graphs}.
\newblock In {\em {SLE}}, volume 7745 of {\em LNCS}, pages 4--23. Springer,
  July 2013.

\bibitem{Voelter2007}
M.~V{\"o}lter and I.~Groher.
\newblock Product line implementation using aspect-oriented and model-driven
  software development.
\newblock In {\em SPLC}, pages 233--242. IEEE, 2007.

\bibitem{Wile1997}
D.~S. Wile.
\newblock {Abstract Syntax from Concrete Syntax}.
\newblock In {\em ICSE}, pages 472--480, 1997.

\bibitem{XingS06a}
Z.~Xing and E.~Stroulia.
\newblock {Refactoring Detection based on UMLDiff Change-Facts Queries}.
\newblock In {\em {WCRE}}, pages 263--274. IEEE, 2006.

\bibitem{Guided2012}
V.~Zaytsev.
\newblock {Guided Grammar Convergence. Full Case Study Report. Generated by
  \texttt{converge::Guided}}.
\newblock {\em {ACM CoRR}}, 6541:1--44, July 2012.

\bibitem{Metasyntactically2012}
V.~Zaytsev.
\newblock {Language Evolution, Metasyntactically}.
\newblock {\em {EC-EASST, BX}}, 49, 2012.

\bibitem{Bidirectionalisation2014}
V.~Zaytsev.
\newblock {Case Studies in Bidirectionalisation}.
\newblock In {\em {Pre-proceedings of TFP'14}}, pages 51--58, May 2014.
\newblock Extended Abstract.

\bibitem{SLEIR2014}
V.~Zaytsev.
\newblock {Software Language Engineering by Intentional Rewriting}.
\newblock {\em {EC-EASST, SQM}}, 65, Mar. 2014.

\bibitem{Parsing2014}
V.~Zaytsev and A.~H. Bagge.
\newblock {Parsing in a Broad Sense}.
\newblock In {\em {MoDELS}}, volume 8767 of {\em LNCS}, pages 50--67. Springer,
  Oct. 2014.

\bibitem{SLPS}
V.~Zaytsev, R.~L{\"a}mmel, T.~van~der Storm, L.~Renggli, R.~Hahn, and
  G.~Wachsmuth.
\newblock {Software Language Processing Suite\footnote{The authors are given
  according to the list of contributors at
  \url{http://github.com/grammarware/slps/graphs/contributors}.}}, 2008--2013.
\newblock \url{http://slps.github.io}.

\end{thebibliography}
}
\end{document}